\documentclass[%
 reprint,
 amsmath,amssymb,
 aps,
]{revtex4-2}

\usepackage{graphicx}
\usepackage{dcolumn}
\usepackage{bm}

\PassOptionsToPackage{numbers, compress}{natbib}

\usepackage{hyperref}
\hypersetup{
    colorlinks=true,
    linkcolor=magenta,
    filecolor=magenta,      
    urlcolor=magenta,
    citecolor=magenta}

\usepackage{geometry} 
\usepackage{url}            
\usepackage{booktabs}       
\usepackage{amsfonts}       
\usepackage{nicefrac}       
\usepackage{microtype}      
\usepackage{xcolor}         
\usepackage{amsmath}
\usepackage{amsthm}
\usepackage{amsfonts}
\usepackage{tcolorbox}
\usepackage{physics}
\usepackage[ruled,linesnumbered,vlined]{algorithm2e}
\usepackage{parskip}
\usepackage{tabularx}
\newcolumntype{C}{>{\centering\arraybackslash}X} 

\usepackage{multirow}
\usepackage{float}

\SetCommentSty{mycommfont}

\SetKwInput{KwInput}{Input}                
\SetKwInput{KwOutput}{Output}

\theoremstyle{definition}
\newtheorem{definition}{Definition}[section]

\theoremstyle{plain}
\newtheorem{lemma}[definition]{Lemma}

\begin{document}

\preprint{APS/123-QED}

\title{On Quantum Natural Policy Gradients}

\author{André Sequeira}
 \email{andresequeira401@gmail.com}
 \affiliation{%
 Departament of Informatics, University of Minho \\
 Quantum Linear and Optical Computation group, International Nanotechnology Laboratory\\
 High Assurance Software Laboratory, INESC TEC\\
 Braga, Portugal
}%
\author{Luis Paulo Santos}%
 \email{psantos@di.uminho.pt}
 \affiliation{%
 Departament of Informatics, University of Minho \\
 Quantum Linear and Optical Computation group, International Nanotechnology Laboratory\\
 High Assurance Software Laboratory, INESC TEC\\
 Braga, Portugal
}%
\author{Luis Soares Barbosa}%
 \email{lsb@di.uminho.pt}
 \affiliation{%
 Departament of Informatics, University of Minho \\
 Quantum Linear and Optical Computation group, International Nanotechnology Laboratory\\
 High Assurance Software Laboratory, INESC TEC\\
 Braga, Portugal
}%
\date{\today}

\begin{abstract}
This research delves into the role of the quantum Fisher Information Matrix (FIM) in enhancing the performance of Parameterized Quantum Circuit (PQC)-based reinforcement learning agents. While previous studies have highlighted the effectiveness of PQC-based policies preconditioned with the quantum FIM in contextual bandits, its impact in broader reinforcement learning contexts, such as Markov Decision Processes, is less clear. Through a detailed analysis of Löwner inequalities between quantum and classical FIMs, this study uncovers the nuanced distinctions and implications of using each type of FIM. Our results indicate that a PQC-based agent using the quantum FIM without additional insights typically incurs a larger approximation error and does not guarantee improved performance compared to the classical FIM. Empirical evaluations in classic control benchmarks suggest even though quantum FIM preconditioning outperforms standard gradient ascent, in general it is not superior to classical FIM preconditioning.\end{abstract}

\maketitle


\section{Introduction}
Reinforcement Learning (RL) emerged as a pivotal technology in modern artificial intelligence, driving progress in diverse fields \cite{sutton_reinforcement_1998, subramanian_reinforcement_2022}. Deep RL, in particular, exceeded human performance in complex tasks, demonstrating its efficacy in games like Atari, Go, and No-limit poker, among others. The integration of RL and Deep Neural Networks (DNNs) placed positioning RL at the forefront of AI for complex sequential tasks in uncertain environments \cite{russell_artificial_2020}. RL's strength lies in its ability to allow software agents to adapt and optimize decision-making in unknown environments. This versatility has led to significant applications in healthcare, autonomous driving, and robotics \cite{arulkumaran_brief_2017}.

In the landscape of RL, the inception of the Natural Policy Gradient (NPG) algorithm \cite{kakade_natural_2001} marks a seminal advancement. This algorithm addresses stability and sample complexity issues -- identified as intrinsic shortcomings of traditional policy gradient methods \cite{sutton_policy_1999,williams_simple_1992}. Specifically, NPG enhances the stability of policy gradient methods by preconditioning the gradient with the inverse of the Fisher Information Matrix (FIM), facilitating updates directly in the policy space and thereby emerging as a highly sample-efficient RL algorithm \cite{agarwal_theory_2021}. However, the efficacy of NPG is still tethered to the curse of dimensionality, exacerbated by the estimation and inversion of the FIM. This limitation catalyzed the evolution of various NPG derivatives, including Trust Region Policy Optimization (TRPO) \cite{schulman_trust_2017} and Proximal Policy Optimization (PPO) \cite{schulman_proximal_2017}, which have been crucial in advancing Deep RL.

Quantum RL models, employing Parameterized Quantum Circuits (PQCs) demonstrated empirically superior sample complexity in addressing fully visible environments compared to a subset of conventional DNNs, as evidenced in standard classical control benchmarking scenarios \cite{skolik_quantum_2022,chen_variational_2020,meyer_quantum_2023,sequeira_policy_2023,jerbi_parametrized_2021}. In \cite{cherrat_quantum_2023}, the authors elucidated that certain PQC-based policies, composed of compound layers, are devoid of barren plateaus, rendering them conducive for financial tasks such as hedging, where deep RL proves instrumental in real market frameworks. Moreover, a quadratic separation in gradient estimation between classical and quantum RL models, provided oracle access to environmental dynamics, was established in \cite{jerbi_quantum_2022}. In \cite{meyer_quantum_2023}, the authors demonstrate empirically that a PQC-based agent doing gradient updates preconditioned by quantum FIM, has better performance compared to standard euclidean updates. Despite these strides, a number of questions remain: Can the sample complexity of PQC-based policies be surely improved by employing quantum natural gradients \cite{stokes_quantum_2020}? What is the actual role of the quantum FIM? This paper aims at contributing to address these questions through exploiting well-known Löwner inequalities \cite{meyer_fisher_2021} between the classical and quantum FIM and its impact in the \emph{regret} of a PQC-based agent. These questions pivots on the potential of quantum NPG as a possible alternative to the classical NPG algorithm, with the prospect of significantly impacting practical applications. This is particular relevant in quantum control \cite{niu_universal_2019}, in which the transition from classical to quantum natural gradients opens a perspective of exploration, potentially harboring enhanced algorithmic stability and sample complexity, thus elevating the robustness and efficiency of RL frameworks.
\subsection*{Related work}
In \cite{meyer_quantum_2023-1} it was empirically demonstrated within the contextual bandits framework that PQC-based policies, performing gradient updates preconditioned by the quantum FIM, exhibit enhanced sample complexity and training stability in comparison to standard Euclidean updates. However, the efficacy of quantum natural policy gradients in broader RL domains beyond contextual bandits, particularly in conventional Markov Decision Processes, remains unexplored. Furthermore, a comprehensive understanding of the quantum FIM's role, as juxtaposed with the classical FIM employed in the original NPG algorithm \cite{kakade_natural_2001}, is yet to be attained. Given the distinct nature of these two information matrices, a pivotal question emerges, which becomes crucial to our investigation:
\\
\\
\textit{Does a PQC-based agent accrue tangible benefits from employing updates in state-space with the quantum FIM as opposed to updates in policy-space with the classical FIM?}

\subsection*{Contributions}
This paper seeks to elucidate the aforementioned query by harnessing well-established Löwner inequalities between the two information matrices \cite{meyer_fisher_2021}. Subsequently, we delineate inequalities concerning the regret of PQC-based agents employing natural gradients preconditioned by both the classical and quantum FIMs. In summary, our main contributions are:
\renewcommand{\labelitemi}{$\ast$}
\begin{itemize}
    \item In the absence of additional insights regarding the nature of the information matrices, a PQC-based agent using the quantum FIM will have a large approximation error compared to the classical FIM and in general not assuring an enhanced regret and thus poorer sample complexity.
    \item If the square root of the information matrices is considered rather than the conventional inverse, the larger approximation error mentioned above could be compensated. However, this does not inherently imply the attainment of the optimal policy.
    \item The performance of PQC-based policies resorting to natural gradients was empirically examined in standard classic control benchmarking environments \cite{sutton_reinforcement_1998}, with gradient preconditioning using 1) the inverse and 2) the square root inverse of the information matrices. It was not observed a substantial improvement when considering the quantum FIM inverse. However, if the square root inverse is employed, the quantum FIM provides an improved sample complexity compared to the square root of classical FIM preconditioning. This indicates that in this setting the matrix compensates for the approximation error.
    \item Sample complexity analysis for the estimation of both quantum and classical FIM, indicates that the quantum FIM is independent of the total number of actions of a given environment, as opposed to the classical FIM. This may be interesting in large action spaces, where samples are expensive to obtain.
\end{itemize}

Section \ref{sec: QPG} provides a comprehensive introduction to policy gradient methods and elaborates on the PQC-based policies under consideration. Section \ref{sec: QNPG} forms the crux of this paper, introducing the QNPG algorithm and discussing key lemmas pertaining to the significance of the quantum FIM in NPG optimization. Section \ref{sec: numerical_experiments} details the experimental framework and shares the findings from these experiments. The paper concludes with Section \ref{sec: conclusion}, where we summarize our findings and explore potential avenues for future research.

\section{Quantum Policy Gradients}\label{sec: QPG}

Policy Gradients aim to learn a parameterized probability distribution over actions given states, a \textit{policy} denoted as $\pi(a\lvert  s,\theta)$, where $\theta \in \mathbb{R}^k$ represents the parameter vector of size $k$, $s \in S$ denotes the state and $a \in A$ the action. The main goal is to perform gradient ascent on a performance metric $J(\theta)$:

\begin{equation}
\theta_{i+1} = \theta_i + \eta \nabla_{\theta_i} J(\theta_i)
\end{equation}

The REINFORCE algorithm \cite{williams_simple_1992} is the simplest policy gradient algorithm, that estimates the gradient of samples obtained from $N$ trajectories of length $T$—also known as the \emph{horizon} - under the parameterized policy, as in Equation \eqref{eq: policy gradient estimator}.

\begin{equation}
    \nabla_{\theta} J(\theta) = \frac{1}{N} \sum_{i=0}^{N-1}\sum_{t=0}^{T-1} (G_t(\tau_i) - b(s_{t_i})) \nabla_{\theta} \log \pi(a_{t_i} \lvert  s_{t_i} , \theta)\label{eq: policy gradient estimator} 
\end{equation}

where $b(s_{t_i})$ is an action-independent control variate also known as \emph{baseline} that is subtracted from the return, resulting in a variance reduction. In this work, the average return was considered as the baseline, computed by Equation \eqref{eq: baseline}.

    \begin{equation} 
    b(s_t) = {1 \over {N}} \sum_{i=0}^{N-1} G_t(\tau_i)
    \label{eq: baseline}
    \end{equation}

In the sequel, the policy $\pi(a|s,\theta)$ shall be regarded as a  PQC-based policy, i.e. the policy is being generated from the output of measurements of PQC's. Specifically, two formulations of such a policy be: the \textit{Born policy} (Definition \ref{def: born policy}) and the \textit{softmax-policy} (Definition \ref{def: softmax policy}).
\vspace{0.5cm}
\begin{definition}
    \label{def: born policy}
    Let $s \in \mathcal{S}$ be a state embedded in an $n$-qubit parameterized quantum state, $\ket{\psi(s,\theta)}$, where $\theta \in \mathbb{R}^k$. The probability associated to a given action $a \in \mathcal{A}$ in the generalized Born framework is given by: 
    \begin{equation}
        \pi(a|s,\theta) = \langle P_a \rangle_{s,\theta} = \bra{\psi(s,\theta)} P_a \ket{\psi(s,\theta)}
    \end{equation}
    where $P_a = \sum_{v \in V_a} \ket{v}\bra{v}$ is the projector into a partition $V_a \subseteq V$ of $|V_a|$ eigenstates of an observable
    \begin{equation}
        O = \sum_{i=0}^{2^n - 1} \lambda_i \ket{i}\bra{i}
    \end{equation}
    Moreover, $\bigcup_{a \in \mathcal{A}} V_a = V$ and $V_a \cap V_{a'} = \emptyset$.
\end{definition}
Definition \ref{def: born policy} presents the most general definition Born policy. However, there could be partitions that do not take into account every eigenstate of a given observable. In these scenarios, the probability associated to a given action would not be normalized as before since $\sum_{a \in A} P_a \neq I$. Moreover, such partitions lead to different local measurement schemes. In the sequel, a contiguous partitioning of the eigenstates of the computational basis measurement will be considered, partitioning into $|\mathcal{A}|$ sets of equal size.
\begin{definition}
    \label{def: softmax policy}
    Let $s \in \mathcal{S}$ be a state embedded in an $n$-qubit parameterized quantum state, $\ket{\psi(s,\theta)}$, where $\theta \in \mathbb{R}^k$. Let $O_a$ be an arbitrary observable representing the numerical preference of action $a \in \mathcal{A}$ and $\beta$ the inverse temperature hyperparameter. The probability associated to a given action $a$ in a softmax policy is given by: 
    \begin{equation}
        \pi(a | s , \theta) = \frac{e^{\beta \langle O_a \rangle_{s,\theta}}}{\sum_{a'} e^{ \beta \langle O_a' \rangle_{s,\theta}}}
    \end{equation}
    $\mathcal{O}(|\mathcal{A}|)$ different observables may be used to attribute the action's numerical preference.
\end{definition}
The policy gradient (Equation \eqref{eq: policy gradient estimator}) is, in its essence, classical with the exception of the log policy gradient in which the gradient w.r.t the PQC must be computed. In that regard, the log policy gradient must be expressed as the gradient of the expectation value of an observable and the parameter-shift rule \cite{schuld_evaluating_2019} can then be applied to compute the gradient using quantum hardware. Let $\langle O \rangle_{\theta}$ be the parameterized expectation value of the observable $O$. The parameter-shift rule is a hardware-friendly technique to compute the partial derivative of $\langle O \rangle_{\theta}$ w.r.t $\theta$. Explicitly, for gates with two eigenvalues, it corresponds to:

\begin{equation}
    \frac{\partial \langle O \rangle_{\theta}}{\partial \theta_l} = \frac{1}{2} \bigl[ \langle O \rangle_{\theta+\frac{\pi}{2} e_l} - \langle O \rangle_{\theta-\frac{\pi}{2} e_l}\bigr]
    \label{eq: parameter-shift rule}
\end{equation}
where $e_l$ indicates that the parameter $\theta_l$ is being shifted. The equality indicates that the partial derivative can be obtained using two quantum circuit evaluations. Thus, for $\theta \in \mathbb{R}^k$, the  gradient can be estimated ideally using $2k$ total quantum circuit evaluations. However, it is known that the expectation value itself can be estimated up to additive error $\mathcal{O}(\epsilon^{-2})$. Thus, $\mathcal{O}(2k\epsilon^{-2})$ quantum circuit calls are needed. For arbitrary functions of expectation values like the log policy gradient, the gradient can be obtained via standard chain rule. For the softmax policy, the log policy gradient takes a peculiar form expressed as a centered version of the gradient of the expectation values encoding the numerical preference of each action \cite{jerbi_parametrized_2021}:
\begin{equation}
    \nabla_{\theta} \log \pi(a|s,\theta) = \beta \biggl[ \nabla_{\theta}\langle O_a \rangle_{\theta} - \mathbb{E}_{a' \sim \pi(\dot|s,\theta)} \nabla_{\theta}\langle O_{a'} \rangle_{\theta} \biggr]
\end{equation}

\section{Natural gradients in policy optimization}\label{sec: QNPG}

This section introduces the QNPG algorithm and delineates its theoretical advantages over the conventional classical NPG. Initially, we discuss the classical NPG algorithm and analyze the regret associated with smooth policies \cite{agarwal_theory_2021}. Subsequently, we propose a reformulation that incorporates the quantum FIM. The derivation of the regret bound for the QNPG algorithm is then grounded in established Löwner inequalities, which compare the classical and quantum FIMs, as detailed in \cite{meyer_fisher_2021}.
   \subsection*{Natural Policy Gradients}
   The Natural Policy Gradient algorithm (NPG) \cite{kakade_natural_2001} is a rescaled version of the policy gradient that performs gradient updates in the geometry induced by the information matrix associated to the policy, the \textit{Fisher Information matrix} (FIM) as follows:
   \begin{equation}
      \label{eq: npg}
      \theta^{t+1} \leftarrow \theta^{t} + \eta F^{-1} \nabla_{\theta} V^{\pi_{\theta}}(\rho)
   \end{equation}
   where $F$ is the average FIM on the sampled states and actions under policy $\pi_{\theta}$ as follows:

      \begin{equation}
         F = \mathbb{E}_{s \sim d^{\pi_{\theta}}} \mathbb{E}_{a \sim \pi_{\theta}(\cdot \mid s)} \bigl[ \nabla_{\theta} \log \pi_{\theta}(a \mid s) \nabla_{\theta} \log \pi_{\theta}(a \mid s)^{T} \bigr]
      \end{equation}
   
   where $d^{\pi_{\theta}}$ is the distribution of states generated under policy $\pi_{\theta}$. Notice that $F$ is positive-definite i.e., $F>0$, however in practice due to instabilities in approximating the information matrix, the inverse $F^{-1}$ is replaced by the Moore-Penrose pseudoinverse $F^{\dagger}$ and regularization is often considered. 
   The notion of \textit{regret} is often considered in RL algorithms as a measure of the difference between the policy being followed and an hypothetical optimal policy. Specifically, regret is computed as the difference between the expected reward of an optimal policy and the reward garnered by the agent's policy over a specified number of episodes or time steps as follows, 
   \begin{equation}
         \sum_{t=1}^{T} \left( V^*(s_t) - V^{\pi}(s_t) \right)
   \end{equation}
where \( V^*(s_t) \) denotes the value function under the optimal policy for state \( s_t \) at time \( t \), and \( V^{\pi}(s_t) \) denotes the value function under the policy \( \pi \) employed by the agent. In \cite{agarwal_theory_2021} the authors established a regret bound for the NPG algorithm considering a general class of smooth parameterized policies. The regret lemma is restated below for completeness.

   \begin{widetext}
   \begin{lemma}[NPG Regret Lemma \cite{agarwal_theory_2021}]
      \label{lemma: regret lemma}   
      Fix a comparison policy $\tilde{\pi}$ and a state distribution $\rho$. Assume for all $s \in \mathcal{S}$ and $a \in \mathcal{A}$ that $\log \pi(a \mid s,\theta)$ is a $\beta$-smooth function of $\theta$. Consider $\pi^{(0)}$ the uniform distribution for every state and the sequence of weights $w^{(0)}, \ldots, w^{(T)}$ satisfying $\left\|w^{(t)}\right\|_2 \leq W$. Let $\epsilon_t$ be the approximation error at time $t$:
      
      \begin{equation}
      \epsilon_t=\mathbb{E}_{s \sim \tilde{d}} \mathbb{E}_{a \sim \widetilde{\pi}(\cdot \mid s)}\left[A^{(t)}(s, a)-w^{(t)} \cdot \nabla_\theta \log \pi^{(t)}(a \mid s)\right]
      \end{equation}
      Then the regret at time step $t$ is upper bounded by:
      \begin{equation}
          \min _{t<T}\left\{V^{\tilde{\pi}}(\rho)-V^{(t)}(\rho)\right\} \leq \frac{1}{1-\gamma}\left(\frac{\log |\mathcal{A}|}{\eta T}+\frac{\eta \beta W^2}{2}+\frac{1}{T} \sum_{t=0}^{T-1} \epsilon_t\right)
      \end{equation}
      \end{lemma}
   \end{widetext}
      where $\tilde{d}$ is the distribution of states generated under the comparison policy $\tilde{\pi}$. $\left\|w^{(t)}\right\|_2$ is the norm of the vector resulting of the multiplication of the inverse of the classical FIM, $F$, by the gradient vector, $w^{(t)} = F^{-1}\nabla_\theta \log \pi^{(t)}(a \mid s)$. $\epsilon_t$ is the approximation error at time step $t$ derived from \textit{compatible function approximation} \cite{sutton_policy_1999}. Lemma \ref{lemma: regret lemma} can thus be utilized in the context of PQC-based policies should these policies respect smoothness conditions. Recall that a function $f: \mathbb{R}^k \mapsto \mathbb{R}$ is $\beta$-smooth if for all $(x,x') \in \mathbb{R}^k$ \cite{agarwal_theory_2021}:  
      \begin{equation}
         \left\| \nabla f(x) - \nabla f(x') \right\|_2 \leq \beta \left\| x - x' \right\|_2
      \end{equation}
      The smoothness of both Born and Softmax policies is established in \cite{jerbi_quantum_2022} through the Gevrey condition. Since $\pi(a|s,\theta) \in [0,1]$ and in the context of RL, where the action is being sampled from the policies probability distribution, it implies that $\pi \in (0,1]$.

   \subsection*{Quantum Natural Policy Gradients}  

   The Quantum Natural Policy Gradient algorithm (QNPG) is obtained by replacing the classical FIM with the QFIM, here represented as $\mathcal{F}$. Restricting ourselves to pure quantum states, the QFIM takes the common form \cite{meyer_fisher_2021}:

   \begin{equation}
      \mathcal{F}_{ij} = 4 \text{Re} \bigl[ \langle \partial_{\theta_i} \psi | \partial_{\theta_j} \psi \rangle - \langle \partial_{\theta_i} \psi |\psi \rangle \langle \psi | \partial_{\theta_j} \psi \rangle \bigr]
   \end{equation}
   In the context of machine learning, a data-dependent QFIM is needed. Therefore, considering again $d^{\pi_{\theta}}$ as the distribution of states generated under parameterized policy $\pi_{\theta}$, the data-dependent QFIM becomes:
   \begin{widetext}
   \begin{equation}
      \label{eq: data-dependent QFIM}
      \mathcal{F}_{ij} = \mathbb{E}_{s \sim d^{\pi_{\theta}}}  4\text{Re} \bigl[ \langle \partial_{\theta_i} \psi(s,\theta) | \partial_{\theta_j} \psi(s,\theta) \rangle - \langle \partial_{\theta_i} \psi(s,\theta) |\psi(s,\theta) \rangle \langle \psi(s,\theta) | \partial_{\theta_j} \psi(s,\theta) \rangle \bigr]
   \end{equation}
\end{widetext}
   Notice that in practice, the empirical QFIM is thus obtained from a finite set of states in a trajectory $T$, obtained under policy $\pi_{\theta}$.
   It is crucial to understand the differences between the FIM and QFIM. Since they are information matrices, they capture what happens in the neighbourhood of a parameter $\theta$ of a given parameterized model by a distance measure. Their difference resorts to what distances are considered within the two different spaces. FIM considers the distance between probability distributions i.e., policies in the context of RL. Thus, the FIM gives information about how the policy changes when infinitesimal changes are performed on a parameter. QFIM, on the other hand, considers distances in the space of quantum states. Thus, it gives information on how the parameterized quantum state changes, given a slight variation of a parameter.

   \subsection*{QFIM as a metric for policy optimization}

   At first glance, one should say that the FIM is more relevant for policy optimization since it captures changes directly in the policy space. However, even though the QFIM is not actually capturing information in the policy space it could be of independent interest since the policy in our case is derived from the quantum state itself. The use of QFIM in policy gradients can be understood as having different impact depending on the type of PQC-based policy employed. For that matter, consider the Softmax policy as presented in Definition \ref{def: softmax policy}. In its most general form it is comprised of $\mathcal{O}(|A|)$ different expectation values encoding numerical preferences. This makes building the connection between QFIM and expectation value of observables a non-trivial and non-intuitive task. On the other hand, the Born policy (Definition \ref{def: born policy}) is derived from projective measurements. Recall that QFIM is derived from the fidelity distance between quantum states \cite{meyer_fisher_2021}. Thus, there is an intricate connection between QFIM and the Born policy. For that reason let us start with the Born policy. 

   Consider a Born policy $\pi(a|s,\theta) = \langle P_a \rangle_{s,\theta}$ in which $P_a$ is the projector into a partition of $V_a$ eigenstates of an observable. For the sake of simplicity, let $V_a$ be a partition of computational basis states and the policy defined as follows:

   \begin{align}
      \pi(a|s,\theta) &= \sum_{v \in V_a} \langle \psi(s,\theta) | v \rangle \langle v | \psi(s,\theta) \rangle \nonumber\\
      &= \sum_{v \in V_a}  | \langle v | \psi(s,\theta) \rangle|^2
   \end{align}

   Recall that QFIM is a metric that describes changes in state space under variation of $\theta$ \cite{haug_generalization_2023} which means that:

   \begin{equation}
      |\langle \psi(s,\theta) | \psi(s,\theta + \delta) \rangle|^2 = 1 - \frac{1}{4} \mathcal{F}_{ij} \delta_i \delta_j 
   \end{equation}

   This has a clear impact on policy optimization since the policy is captured in the same way as projectors onto a partition of basis states. More importantly recall that classical FIM corresponds to the information matrix associated to the probability distribution generated from the measurement of the quantum state where we could say that the measurement $\mathcal{M} = \{ P_a \}$ where $P_a$ is the ${a^\text{th}}$ outcome of the experiment , corresponds to the partition of action $a$. In this setting, the following matrix inequality \cite{meyer_fisher_2021} applies:

   \begin{equation}
      F \leq \mathcal{F}
      \label{eq: lowner ineq fisher}
   \end{equation}

   Inequality \eqref{eq: lowner ineq fisher} expresses the Löwner inequality of positive semi-definite matrices \cite{bhatia_matrix_1997} i.e., $\mathcal{F} - F \geq 0$ has only non-negative eigenvalues. The inequality indicates that QFIM is always an upper bound for any information matrix obtained from the outcome of measurements in a parameterized quantum state. The equality happens once the parameterized quantum state prepares a classical probability distribution. The matrix inequality forms the basis for the separation in terms of agent's regret presented in this work. 

   The NPG objective is optimizing the policy under the log policy gradient, which is slightly different compared to standard quantum natural gradient objective. Nevertheless, recall that natural gradients indeed perform gradient updates using adaptive step sizes, and for that matter if we expand the log policy gradient as $\nabla \text{log}(\pi) = \frac{\nabla \pi}{\pi}$, we could embed the resulting denominator into an adaptive learning rate $\eta' = \frac{\eta}{\pi}$, resulting in approximately the same objective as in standard quantum natural gradient. Thus, for the Born policy QFIM has a direct impact from the state space to the policy space. 
   
   For the Softmax policy though the situation is not so intuitive. Recall that to build the policy requires estimating the expectation values of $\mathcal{O}(|A|)$ observables. Therefore, Inequality \eqref{eq: lowner ineq fisher} lost meaning in this scenario since the classical FIM would not be generated from the output of a fixed quantum measurement but from a distribution obtained from $\mathcal{O}(|A|)$ possibly different expectation values. However, recall that the softmax policy was originally considered to overcome the lack of greediness control in the Born policy \cite{jerbi_parametrized_2021}. That is, at time step $T$ for some environment we could already know everything about the reward function but nonetheless, the type of parametrization could for instance not allow for deterministic policies. For that reason, the Softmax policy is usually considered where an hyperparameter $\beta$ control its greediness. In this setting, instead of considering the most general softmax formulation and have $\mathcal{O}(|A|)$ expectation values of operators in multiple bases, the observable could then simply be the projectors considered in the Born policy, i.e. $\langle O_a \rangle = \langle P_a \rangle$. In this scenario, expanding the log policy gradient leads to the NPG gradient update for the Softmax policy: 

   \begin{widetext}
   \begin{equation}
      \eta\mathcal{F}^{-1} \nabla_{\theta} \log \pi_{\theta}(a | s,\theta) =  \beta \biggl[\eta\mathcal{F}^{-1} \nabla_{\theta} \langle P_a \rangle - \mathbb{E}_{a \sim \pi(.|s,\theta)}[ \eta\mathcal{F}^{-1} \nabla_{\theta} \langle P_{a'} \rangle ] \biggr]
   \end{equation}
\end{widetext}

   where $P_a$ is the projector into a partition of basis states $V_a$ as before and $\mathbb{E}_{a \sim \pi(.|s,\theta)}$ the expectation under the policy. QFIM would then have the same impact in policy optimization, however taking into consideration every action as opposed to the Born policy, and modifying the update to take a centered version of the natural gradient into account. Even though Inequality \eqref{eq: lowner ineq fisher} would not in principle apply in this scenario, we would expect such gradient update to be beneficial nonetheless in policy optimization. It remains to be seen in practice the actual role of the QFIM in policy optimization under the Softmax policy.

   \subsection*{QFIM for improved regret}

   The NPG regret Lemma \ref{lemma: regret lemma} establishes that the regret of agent that uses an arbitrary and smooth parameterized policy is dependent on the vector norm $||w||_{2}$ and the compatible function approximation error $\epsilon_t$. Thus, to establish bounds on the regret dependently on the information matrix employed, it would suffice to establish bounds on the norms and the approximation errors presented in the regret lemma, induced by those information matrices. Let us start with the norms. Let $||w_{\mathcal{F}}||_2$ and $||w_F||_2$ be the 2-norm induced by QFIM and classical FIM, respectively.
   The goal of this section is to clarify in which conditions we have the norm inequality
   \begin{equation}
      ||w_{\mathcal{F}}||_2 \leq ||w_F||_2
      \label{eq: desired norm inequality}
   \end{equation}

   Thus indicating that the regret associated to PQC-based agent employing NPG optimization benefits from considering the QFIM as the metric instead of the classical FIM.
   
   Let $F$ and $\mathcal{F}$ be two positive semi-definite matrices such that $F \leq \mathcal{F}$ i.e., $\mathcal{F} - F \geq 0$ has only non-negative eigenvalues. Let $v = \nabla_{\theta} \log \pi_{\theta}(a|s,\theta)$. Let $||w_{F}||_2 = ||F^{-1} v||_2$ and $||w_{\mathcal{F}}||_2 = ||\mathcal{F}^{-1} v||_2$ be 2-norm induced by the FIM and QFIM, respectively. Thus:
   \begin{equation}
      F \leq \mathcal{F} \quad \not\Rightarrow \quad ||w_{\mathcal{F}}||_2 \leq ||w_F||_2
   \end{equation}
   for all $v \in \mathbb{R}^k$. That is, the matrix inequality does not readily imply the vector norm inequality for every gradient vector. Moreover, notice that we are considering positive semi-definite matrices, but the inequality actually considers the inverses and not the pseudoinverses. However, in practice, both QFIM and FIM are ill-conditioned and thus they need to be regularized before inversion i.e., $\mathcal{F} = \mathcal{F} + \epsilon I$ where $I$ is the identity and $\epsilon > 0$ is the regularization term. For that reason, let us consider the inverses from now on. The Löwner partial order inequality guarantees the reverse inequality for the inverses of positive (semi-)definite matrices.

   \begin{equation}
      F \leq \mathcal{F} \quad \text{iff} \quad F^{-1} \geq \mathcal{F}^{-1}
      \label{eq: lowner partial order psd matrices}
   \end{equation}

   Thus, the inequalities \eqref{eq: lowner partial order psd matrices} can be used to establish the conditions for which the desired vector norm inequality in Equation \eqref{eq: desired norm inequality} is reached. 

   From the definition of positive semi-definite matrices we have that for any vector $v \in \mathbb{R}^k$ the following applies:
   \begin{equation}
      v^T F v \geq 0 \quad \text{and} \quad v^T \mathcal{F} v \geq 0
   \end{equation}
   which implies that
   \begin{equation}
      F \leq \mathcal{F} \quad \implies \left\{
         \begin{array}{ll}
               v^T F v \leq v^T \mathcal{F} v \\
               v^T F^{-1} v \geq v^T \mathcal{F}^{-1} v\\
         \end{array} 
         \right. 
   \end{equation}
   Recall that 2-norm $|| F v ||_2^2 = \biggl( F v \biggr)^{T} \biggl( F v\biggr)$ and since in this case both information matrices are Hermitian $(F=F^{T})$ then, 
   \begin{align}
      || F v ||_2^2 &= \biggl( F v \biggr)^{T} \biggl( F v\biggr) = v^{T} F^{T} F v = v^{T} F^2 v 
      \label{eq: expansion of 2-norm}
   \end{align}
   since $F=F^{T}$. Thus, the vector norm inequality that we have been seeking implies the matrix norm inequality
   \begin{align}
      || F v ||_2^2 \quad \implies \quad F^2 \leq \mathcal{F}^2
   \end{align}
   which is not guaranteed in general if the Löwner inequality $F \leq \mathcal{F}$ is all we have. That is, in general
   \begin{equation}
      F \leq \mathcal{F} \quad \not\Rightarrow \quad F^2 \leq \mathcal{F}^2
   \end{equation}
   The implication would be guaranteed if either $F$ or $\mathcal{F}$ is idempotent i.e., $\{0,1\}$ would be their only eigenvalues and the only non-singular matrix (full-rank) would be the identity. This restricts the set of information matrices and thus the set of PQCs needed for the vector norm inequality to be guaranteed. For instance, the PQC
   \begin{equation}
      |\psi(\theta)\rangle = \bigotimes_{i=1}^{n} \cos(\theta_i) |0\rangle + \sin(\theta_i) |1\rangle
   \end{equation}
   has $\mathcal{F} = I$ which would apply. However, it is also true that in this case $F=\mathcal{F}$ thus entailing equality. Therefore, the desired norm inequality would not be in general guaranteed just from the matrix inequality of information matrices. However, notice that the expansion in Equation \eqref{eq: expansion of 2-norm} can also be taken into account considering $F^{\frac{1}{2}}$ instead of $F$. Thus,
   \begin{equation}
      || F^{\frac{1}{2}} v ||_2^2 = v\biggl( F^{\frac{1}{2}} v \biggr)^{T} \biggl( F^{\frac{1}{2}} v\biggr) = v^{T} {F^{\frac{1}{2}}}^{T} F^{\frac{1}{2}} v = v^{T} F v
   \end{equation}
   and since $ v^{T} F v \leq  v^{T} \mathcal{F} v$ and  $v^{T} F^{-1} v \geq  v^{T} \mathcal{F}^{-1} v$ the vector norm inequality is guaranteed:
   \begin{equation}
      || F^{-\frac{1}{2}} v ||_2^2 \geq || \mathcal{F}^{-\frac{1}{2}} v ||_2^2 \quad \Longleftrightarrow \quad F \leq \mathcal{F}
   \end{equation}
   Therefore, a norm inequality depends on the type of information matrix inverse considered. In summary:
   \begin{itemize}
      \item \textbf{$(F^{-1}, \mathcal{F}^{-1})$} - If the standard inverses are considered then the norm inequality is not in general guaranteed, since $F \leq \mathcal{F} \quad \not\Rightarrow \quad ||w_{\mathcal{F}}||_2 \leq ||w_F||_2$, and further information about these matrices is needed.
      \item \textbf{$(F^{-\frac{1}{2}}, \mathcal{F}^{-\frac{1}{2}})$} - Norm inequality is guaranteed since $|| F^{-\frac{1}{2}} v ||_2^2 \geq || \mathcal{F}^{-\frac{1}{2}} v ||_2^2 \quad \Longleftrightarrow \quad F \leq \mathcal{F}$. However its actual utility in solving a RL problem is unknown.
   \end{itemize}
   This result motivates the use of a Generalized Quantum Natural Policy Gradient (GQNPG) algorithm, which for $\varphi \in [0,1]$, the GNQPG algorithm performs the following update: 
   \begin{equation}
      \theta^{t+1} \leftarrow \theta^{t} + \eta \mathcal{F}^{-\varphi} \nabla_{\theta} V^{\pi_{\theta}}(\rho)
      \label{eq: GQNPG}
   \end{equation}

   In \cite{haug_optimal_2021}, the authors suggest a similar update for the standard gradient ascent considering QFIM as metric. The authors suggest that $\varphi=\frac{1}{2}$ constitutes a intricate optimization strategy. As previously described, the standard QFIM is usually ill-conditioned and requires to be regularized $\mathcal{F} = \mathcal{F} + \epsilon I$ where $\epsilon > 0$ could have a dramatic impact on sensitivity to parameter updates and lead to an increase in gradient steps to achieve the convergence of the algorithm. The authors show that, for $\varphi=\frac{1}{2}$, QFIM is intrinsically regularized and thus it is full-rank and does not need $\epsilon$, once the fidelity cost-function is considered. The authors observed that for several PQCs the infidelity had a sharp increase for $\varphi \geq 0.6$ due to ill-conditioned QFIM. They suggest however that for small infidelities standard QFIM with $\epsilon=0.1$ may perform better. However, in the context of policy gradients, it may be very well the case that appears in the beginning of training large infidelities i.e., the policy being far from the optimal policy are expected. Thus, the role of $\varphi$ and the tradeoff between regularization and performance in the context of RL agents should also be addressed besides the standard preconditioning considered in the NPG algorithm.\\

   The approximation error in the regret lemma of Section \ref{sec: QNPG} depends on the type of information matrix employed. Same as before, the inequality between the classical and quantum information matrices imply an inequality between the approximation errors induced by these matrices. Recall that the approximation error at time step $t$, $\epsilon_t$ is defined as: 

   \begin{equation}
      \epsilon_t=\mathbb{E}_{s \sim \tilde{d}} \mathbb{E}_{a \sim \widetilde{\pi}(\cdot \mid s)}\left[A^{(t)}(s, a)-w^{(t)} \cdot \nabla_\theta \log \pi^{(t)}(a \mid s)\right]
    \end{equation}

   For simplicity, let $v = \nabla_\theta \log \pi^{(t)}(a \mid s)$ and $w^{(t)}$ be expanded as a function of the type of information matrix as before. Let $\epsilon_{F}$ and $\epsilon_{\mathcal{F}}$ be the approximation errors induced by the classical and quantum FIMs, respectively. Consider the difference between the approximation errors induced by the classical and quantum FIM,
   \begin{align}
      \epsilon_{\mathcal{F}} - \epsilon_{F} &= - w_{\mathcal{F}} \cdot v + w_{F} \cdot v \nonumber\\
                                             &= - \mathcal{F}^{-1} v \cdot v + F^{-1} v \cdot v \nonumber\\
                                             &= - v^{T} \mathcal{F}^{-1} v +  v^{T} F^{-1} v \nonumber\\
                                             &= v^{T} (F^{-1} - \mathcal{F}^{-1}) v \geq 0 
   \end{align}

   which implies that the approximation error under quantum FIM will always be greater than or equal to the approximation error under classical FIM:
   \begin{equation}
      \epsilon_{\mathcal{F}} \geq \epsilon_{F}
   \end{equation}
   Therefore, for an agent employing the classical FIM for precondition the gradient will have a regret less than or equal to the regret of an agent employing the quantum FIM since the norm inequality is not guaranteed and the quantum FIM actually provides a greater approximation error. However, recall that the Löwner-Heinz inequality \cite{zhan_1_2002} implies that:
   \begin{equation}
      I \leq \mathcal{F} \quad \implies \quad I^{-\frac{1}{2}} \geq \mathcal{F}^{-\frac{1}{2}} 
   \end{equation}
   since for any $0 \leq r \leq 1$, $I^r \leq \mathcal{F}^r$. The approximation error of the square root of classical FIM is then also less than or equal to the square root of quantum FIM,
   \begin{equation}
      \epsilon_{\mathcal{F}^{\frac{1}{2}}} \geq \epsilon_{F^{\frac{1}{2}}}
   \end{equation}
   Therefore, even though the approximation error persists, the regret can be compensated by the norm inequality using the square root of the information matrices. It remains to see in practice now, if the approximation error increases due to quantum FIM can actually be compensated by the norm inequality, since this depends heavily on the problem at hand. The results are summarized in Table \ref{tab: summary results}.

   \begin{table}[!h]
      \begin{tabular}{|c|c|c|c|}
      \hline
      \textbf{$F / \mathcal{F}$}                               & \textbf{$||w_{\mathcal{F}}||_2 \leq ||w_{F}||_2$} & \textbf{$\epsilon_{\mathcal{F}} \leq \epsilon_{F}$} & \textbf{Improved regret} \\ \hline
      \textbf{$F^{-1} / \mathcal{F}^{-1}$}                     & No                                                & No                                                  & No                       \\ \hline
      \textbf{$F^{-\frac{1}{2}} / \mathcal{F}^{-\frac{1}{2}}$} & Yes                                               & No                                                  & ?                        \\ \hline
      \end{tabular}
      \caption{Summary of results. The first column indicates the type of information matrix considered. The second and third columns indicate wether the norm and approximation error inequalities are guaranteed, respectively. The fourth column indicates if the regret is improved. \label{tab: summary results}}
      \end{table}

\section{Performance Evaluation in Benchmarking Environments}\label{sec: numerical_experiments}

In this section, we assess the efficacy of the GQNPG algorithm, as introduced in Section \ref{sec: QNPG}, using two classical control benchmarking environments \cite{sutton_reinforcement_1998}. We selected the Cartpole and Acrobot environments due to their compact state-action spaces, which have previously been efficiently addressed using PQC-based policies \cite{jerbi_parametrized_2021}.

\begin{figure}[!htb]
\centering
\includegraphics[width=0.5\textwidth]{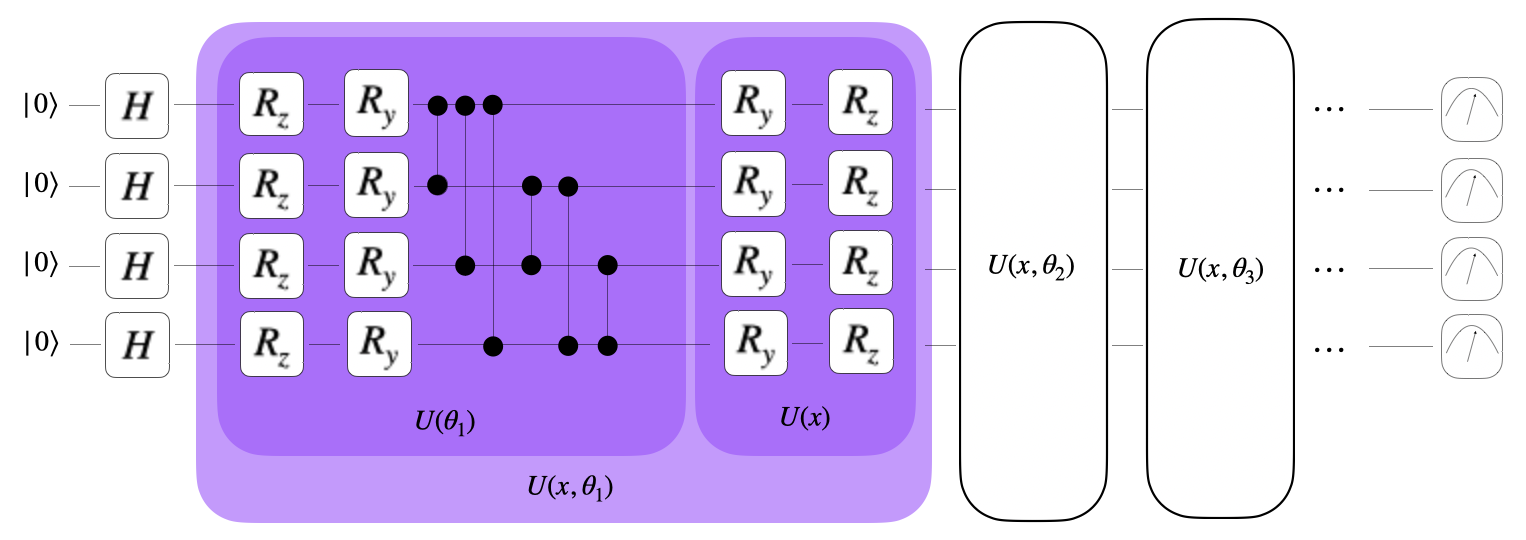}
\caption{The parameterized quantum circuit used in the numerical experiments. Data reuploading is consistent with \cite{jerbi_parametrized_2021}, but input scaling was excluded to improve the estimation of the Quantum FIM matrices.}
\label{fig:PQC}
\end{figure}

\begin{figure}[!htb]
\centering
\includegraphics[width=0.5\textwidth]{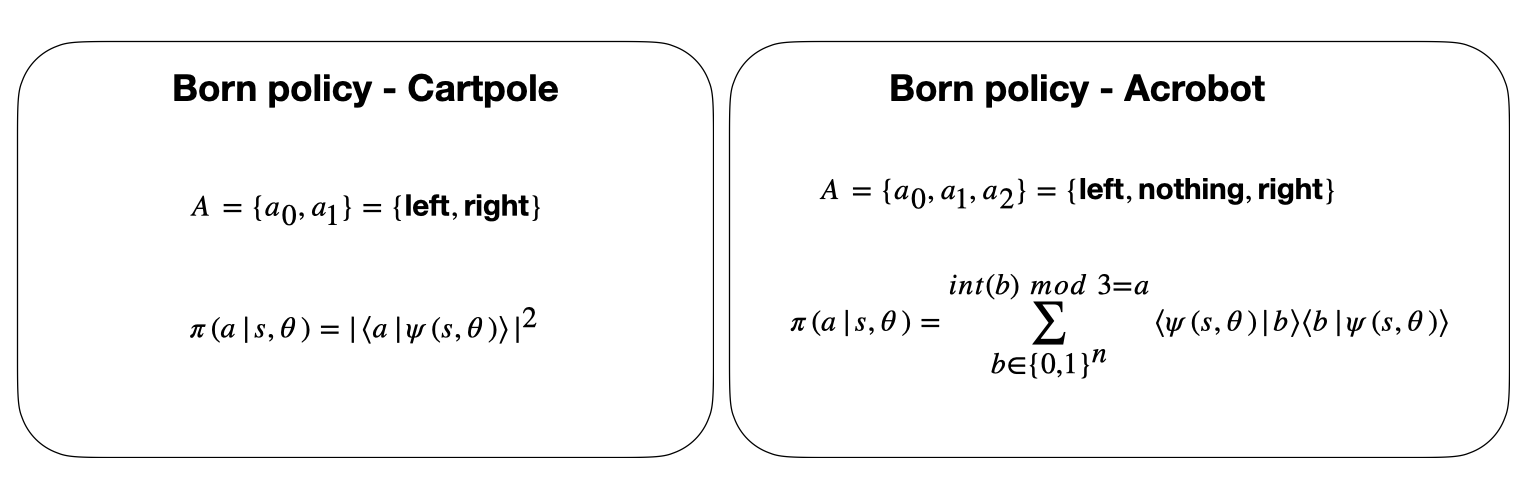}
\caption{Born policies for Cartpole and Acrobot environments.}
\label{fig:born_policies}
\end{figure}

The Cartpole environment features a four-dimensional state space with two potential actions, while the Acrobot environment has a six-dimensional state with three available actions. Notably, in the Acrobot environment, four of the features represent the sine and cosine values of the two joint angles. To optimize training time and reduce the PQC size, we limited the state representation to the angles, thus reducing it to four features. Consequently, both environments utilize the PQC depicted in Figure \ref{fig:PQC}, as proposed by \cite{jerbi_parametrized_2021}, albeit with different layer configurations and measurement strategies. A comprehensive characterization of the environment and the PQC configurations can be found in Table \ref{table: envs} and Table \ref{table: pqcs}, respectively.
We investigated both Born and Softmax PQC-based policies as discussed in Section \ref{sec: QPG}. Simple computational basis measurements were employed to link quantum measurements to their respective policies. For the Cartpole, a single-qubit projector was used. The probability distribution over basis states was estimated to establish the Born policy. In contrast, for the Acrobot, a $\text{mod}-3$ Born policy was adopted as in \cite{jerbi_parametrized_2021}. In this case, each qubit is measured in the computational basis, and a basis state \( b \) is associated with action \( a \) if \( \text{int}(b) \mod 3 = a \). The Born policies for both environments are illustrated in Figure \ref{fig:born_policies}.
For the Softmax policy, while the same projectors as in the Born policy were employed, the probability serves as a numerical preference for a specific action. This preference is subsequently processed by the softmax function to yield a probability distribution over actions. It is important to note that the Softmax policy introduces an inverse temperature hyperparameter, \( \beta \), which influences the policy's greediness, a feature absent in the Born policy. The optimal \( \beta \) value is environment-specific and typically identified through hyperparameter tuning. In our study, we adopted a linear annealing schedule for \( \beta \), starting at 1 and culminating in the final \( \beta \) value as suggested in \cite{jerbi_parametrized_2021}.

\begin{figure*}[!htb]
    \centering
    \includegraphics[width=\textwidth]{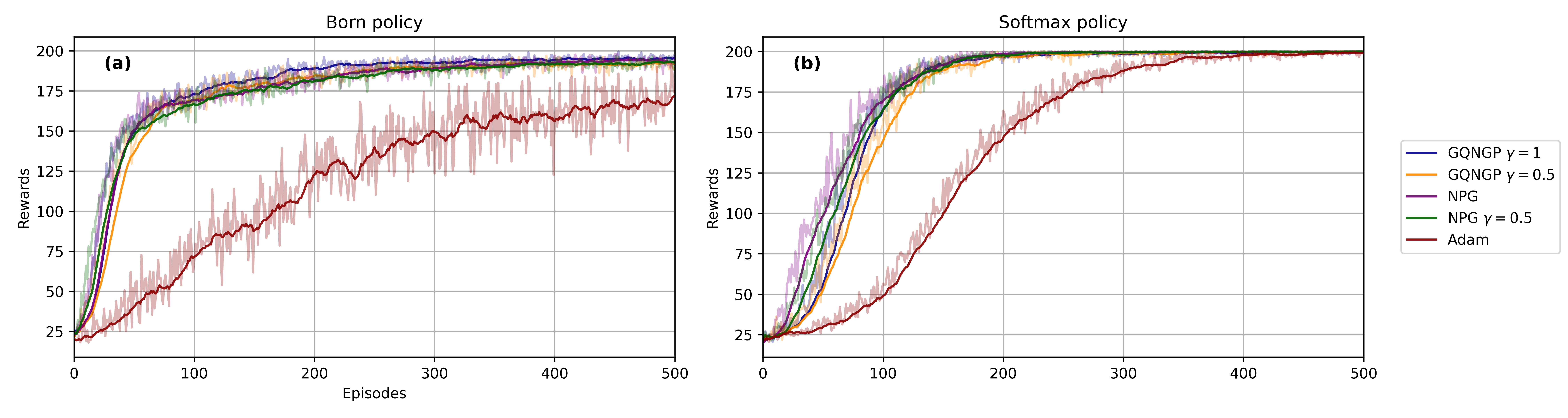}
    \caption{Performance of the NPG algorithm (and its generalized quantum counterpart) in the Cartpole environment. Subfigures \textbf{(a)} and \textbf{(b)} represent the performance of Born and Softmax policies using the cumulative reward as the evaluation metric.}
    \label{fig:gnpg_cartpole}
\end{figure*}

\begin{figure*}[!htb]
    \centering
    \includegraphics[width=\textwidth]{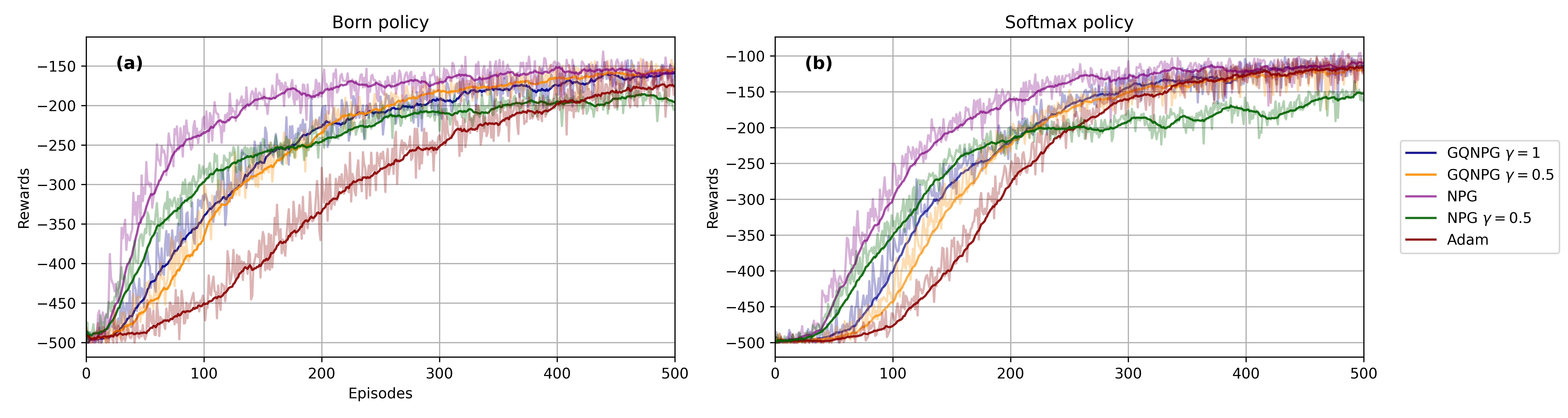}
    \caption{Performance of the NPG algorithm (and its generalized quantum counterpart) in the Acrobot environment. Subfigures \textbf{(a)} and \textbf{(b)} showcase the performance of Born and Softmax policies using cumulative reward as a performance measure.}
    \label{fig:gnpg_acrobot}
\end{figure*}

Performance outcomes for five different optimizers in the Cartpole and Acrobot environments are depicted in Figures \ref{fig:gnpg_cartpole} and \ref{fig:gnpg_acrobot} respectively. The following optimizers were considered:

\begin{itemize}
\item \textbf{Adam}: Utilizes the standard Adam optimizer with a learning rate of \( 10^{-2} \).
\item \textbf{NPG}: Employs the standard NPG algorithm with classical FIM.
\item \textbf{NPG \( \varphi = 0.5 \)}: Uses the NPG algorithm with the square root of the classical FIM.
\item \textbf{GQNPG}: Integrates the NPG algorithm with quantum FIM.
\item \textbf{GQNPG \( \varphi = 0.5 \)}: Adopts the NPG algorithm with the square root of the quantum FIM.
\end{itemize}

The optimizers' performances were benchmarked using the cumulative reward metric, plotted on the y-axis, against the total episode count on the x-axis. Each optimizer's performance was averaged across 50 trials, with each figure displaying a 10-episode running mean and a shaded region representing the standard deviation of the experiments.

Given the deterministic nature of the environments, actions consistently lead to the same observed states and rewards. Furthermore, in accordance with the NPG regret lemma, we employed a zero-initialization approach concerning the parameters of the PQC to effectively have an uniform policy at the beginning of training. This means every parameter in the PQC illustrated in Figure \ref{fig:PQC} was initialized at zero, and since the employed PQC is composed by an initial chain of Hadamard gates, it ensures that policy is in fact uniform and moreover, the variance of the algorithm could solely be attributed to the agent's sampled trajectories.

Our experiments utilized Pennylane's quantum simulator \cite{bergholm_pennylane_2022} with PyTorch-based automatic differentiation. For replication purposes, our work can be accessed through the following GitHub repository \href{https://github.com/andre-sequeira10/GQNPG}{GQNPG}.

A direct comparison between the Born and Softmax policies for the Cartpole environment is available in Figures \ref{fig:gnpg_cartpole}\textbf{(a)} and \ref{fig:gnpg_cartpole}\textbf{(b)}. Notably, the Softmax agents exhibit superior and more consistent performance compared to their Born counterparts. This advantage is attributed to $\beta$ and the ability to regulate the policy's greediness, an ability the Born policy lacks \cite{jerbi_parametrized_2021}. Both policies demonstrate negligible performance variation across different optimizers. However, slight advantages for the GQNPG algorithm in the Born policy could be observed, although these may be a result of statistical variances. A key observation is that gradient preconditioning, regardless of using quantum or classical FIMs, yields similar results. Such result indicate that in this context, updates in state space could be as effective as updates in policy space. That is, the quantum FIM obtained from infinitesimal distances between quantum states is as efficient as the classical FIM which is obtained from infinitesimal distances between policies directly. 

Figures \ref{fig:gnpg_acrobot}\textbf{(a)} and \ref{fig:gnpg_acrobot}\textbf{(b)} depict the performance registered in the Acrobot environment for the Born and Softmax policies, respectively. The Acrobot environment with three actions and a slightly more complex reward function becomes a more complex environment to be solved compared to Cartpole. Such complexity difference implies a more clear separation in optimizer performance compared to the Cartpole. It can be immediately observed that in this case, for both policies, not every variant of natural optimizers performed better than the standard Adam. However, there is a clearer separation between the performance associated to the classical NPG and GQNPG optimizers. In this setting, the classical NPG has more evidently better convergence even though for this environment there is not a clear condition in which the environment is considered solved. Thus, the asymptotic behavior is used here to attribute that the classical NPG algorithm necessitates slightly fewer episodes to reach an asymptote in the cumulative reward. Moreover, both optimizers seem to agree in the same policy after 500 episodes. Furthermore, it is more clear as well that in the Born policy, the GQNPG algorithm with $\varphi=0.5$ performs better than the NPG algorithm with $\varphi=0.5$. In this setting, however, the same conclusion can be reached for the Softmax policy even though the matrix inequalities can not be guaranteed, observed as before. It is curious to observe that in this scenario, the unregularized NPG with $\varphi=0.5$ after a great learning period of around 200 episodes seems to saturate and perform worse than the Adam optimizer.

The results obtained experimentally shined a light at the need to test PQC-based policies with different natural optimizers in even more complex environments characterized by multiple state-action spaces and reward functions to be able to further conclude about the efficacy of quantum FIM based natural policy gradient algorithms.

\section{Comparative analysis for the estimation of information matrices}\label{sec: comparative_analysis}

In this section, we draw a comparison in terms of the resources needed to compute quantum and classical FIM's. The chosen metric to characterize the resources is the number of quantum measurements or quantum circuit executions required to estimate the information matrices. This way, a sample complexity analysis can be made and a possible separation between the two natural gradients assessed. Sample complexity in this context has a specific meaning. It corresponds to the total number of quantum circuit executions and not to the total number of episodes needed to solve an environment, as in standard RL notation. 

\subsection{Sample complexity of estimating classical FIM}
Recall that the classical FIM is represented as the outer product of the gradient of the log policy averaged through the sampled trajectories,

\begin{equation}
    F = \mathbb{E}_{s \sim d^{\pi_{\theta}}} \mathbb{E}_{a \sim \pi_{\theta}(\cdot \mid s)} \bigl[ \nabla_{\theta} \log \pi_{\theta}(a \mid s) \nabla_{\theta} \log \pi_{\theta}(a \mid s)^{T} \bigr]
    \label{eq: FIM}
\end{equation}

Since the gradient of log policy is needed, the sample complexity is actually dependent on the type of policy employed. Let us start discussion with the Born policy. 

\subsection*{FIM - Born policy}

The Born policy is represented as a probability distribution over a partition $V_a$ of computational basis states, as presented in Section \ref{sec: QNPG}. Notice, however, that complexity depends on this partition. In its most general form we could $\pi(a|s,\theta) = \langle P_a \rangle_{s,\theta} = \sum_{v \in V_a} \langle P_v \rangle_{s,\theta} $ with $V_a = \frac{2^n}{|A|}$. This is similar to the representation of the Born policy employed in the Acrobot environment in Section \ref{sec: numerical_experiments}, with the exception that the number of actions is not even and the partition does not perfectly correspond to $\frac{2^n}{|A|}$. Nevertheless, note that in the Cartpole environment the policy is even more simple than before since a single-qubit is considered. The log policy gradient can be expanded in this case using parameter-shift rules, as follows,

\begin{align}
    \partial_{\theta} \log \pi(a|s,\theta) &= \sum_{v \in V_a} \frac{\partial_{\theta} \langle P_v \rangle_{s,\theta}}{\langle P_v \rangle_{s,\theta}} \nonumber \\
    &= \sum_{v \in V_a} \frac{\langle P_v \rangle_{s,\theta + \frac{\pi}{2}} - \langle P_v \rangle_{s,\theta - \frac{\pi}{2}}}{\langle P_v \rangle_{s,\theta}}
\end{align}
It is known that for an $\epsilon$-approximation to the probability $\mathcal{O}(\epsilon^{-2})$ circuit executions are needed. Ignoring the approximation error, the total number of independent quantum circuits needed to estimate the gradient is 3. Each projector is a linear expectation value depending on $\frac{2^n}{|A|}$ partitions.Since FIM is a $k \times k$ matrix for $\theta \in \mathbb{R}^k$, we need $\mathcal{O}(3\frac{2^n}{|A|}k^2)$ quantum circuit executions. 
\subsection*{FIM - Softmax policy}
Assume, for simplicity, that the same projectors as in the Born policy are considered as action's numerical preferences, but are otherwise irrelevant. Recall the log policy gradient expansion:

\begin{equation}
    \nabla_{\theta} \log \pi(a|s,\theta) = \nabla_{\theta}\langle P_a \rangle_{\theta} - \mathbb{E}_{a' \sim \pi(\dot|s,\theta)} \nabla_{\theta}\langle P_a' \rangle_{\theta}
\end{equation}
Thus, the Softmax policy depends on the total number of actions $|A|$ to estimate the derivative w.r.t a single parameter. Thus, using parameter-shift rules for estimating the partial derivatives of projectors as above, for $\theta \in \mathbb{R}^k$, we need $\mathcal{O}(4|A|k^2)$ quantum circuit executions.

\subsection{Sample complexity of estimating quantum FIM}

Recall that the quantum FIM obtained from the infinitesimal distances between quantum states as represented in Equation \eqref{eq: data-dependent QFIM} depends on the quantum state only. Thus, it can be immediately concluded that the sample complexity of estimating quantum FIM will not be dependent on the policy and thus on the total number of possible actions associated with an environment. Importantly, an entry of the quantum FIM, $\mathcal{F}_{ij}$ can be obtained from the estimation of four independent overlaps, as proposed in \cite{meyer_fisher_2021} shifting the respective parameters $i,j$:

\begin{equation}
    \begin{array}{r}
    \mathcal{F}_{i j}=-\frac{1}{2}\left(\left|\left\langle\psi(\boldsymbol{\theta}) \mid \psi\left(\boldsymbol{\theta}+\left(\boldsymbol{e}_i+\boldsymbol{e}_j\right) \frac{\pi}{2}\right)\right\rangle\right|^2\right. \\
    -\left|\left\langle\psi(\boldsymbol{\theta}) \mid \psi\left(\boldsymbol{\theta}+\left(\boldsymbol{e}_i-\boldsymbol{e}_j\right) \frac{\pi}{2}\right)\right\rangle\right|^2 \\
    -\left|\left\langle\psi(\boldsymbol{\theta}) \mid \psi\left(\boldsymbol{\theta}-\left(\boldsymbol{e}_i-\boldsymbol{e}_j\right) \frac{\pi}{2}\right)\right\rangle\right|^2 \\
    \left.+\left|\left\langle\psi(\boldsymbol{\theta}) \mid \psi\left(\boldsymbol{\theta}-\left(\boldsymbol{e}_i+\boldsymbol{e}_j\right) \frac{\pi}{2}\right)\right\rangle\right|^2\right)
    \end{array}
\end{equation}

where $e_j$ is the unit vector along the $\theta_j$ axis. Thus, for $\theta \in \mathbb{R}^k$, we need $\mathcal{O}(4k^2)$ quantum circuit executions to estimate the quantum FIM. 
In conclusion, it seems that the estimation of the quantum FIM may be significantly cheaper compared to that of the classical FIM, especially in the context of using a Softmax policy since every possible action must be taken into account to estimate the classical FIM. However, since the quantum FIM produces updates directly in state-space instead of policy-space, such a difference in sample complexity can be neglected in terms of the actual ability in solving the environment as discussed in Section \ref{sec: numerical_experiments}.

\section{Conclusion}\label{sec: conclusion}

In this paper, we reported a series of experiments aiming at comparing the effectiveness of natural policy gradients preconditioned by the quantum Fisher Information Matrix (FIM) with those preconditioned by the traditional classical FIM. Our findings indicate that considering a quantum FIM preconditioning leads to a larger approximation error. However, when utilizing the square roots of the information matrices, the square root of the quantum FIM could compensate the approximation error with the gradient vector norm which leads to a reduction in regret relative to its classical counterpart. Note however, that this advantage may not always translate into near-optimal policy. This hypothesis was tested in standard control benchmark settings, confirming that the preconditioning of the quantum FIM with its square root inverse leads to better sample efficiency over the square root of the classical FIM preconditioning. Conversely, using the full inverse for quantum FIM preconditioning did not significantly outperform the classical approach. It is important to note that our sample complexity analysis revealed that unlike the classical FIM, the quantum FIM's estimation is not affected by the size of the action space in a given environment, which presents a notable distinction between the two. Further investigation is necessary, particularly in environments with large action spaces since these are not easily solved with current quantum technologies, to fully determine the practical efficacy of quantum natural policy gradients. This will be a focus of future research, along with the investigation of approximations of quantum FIM \cite{beckey_variational_2022,stokes_quantum_2020}. The role of the quantum and classical FIM in the trainability of PQC-based policies is also a promising avenue for future research. 

\subsection*{Acknowledgements}
This work is financed by National Funds through the Portuguese funding agency, FCT - Fundação para a Ciência e a Tecnologia, within grants LA/P/0063/2020, UI/BD/152698/2022 and project IBEX, with reference PTDC/CC1-COM/4280/2021

\bibliographystyle{plainnat}
\bibliography{QNPG}

\begin{thebibliography}{28}
\providecommand{\natexlab}[1]{#1}
\providecommand{\url}[1]{\texttt{#1}}
\expandafter\ifx\csname urlstyle\endcsname\relax
  \providecommand{\doi}[1]{doi: #1}\else
  \providecommand{\doi}{doi: \begingroup \urlstyle{rm}\Url}\fi

\bibitem[Agarwal et~al.(2021)Agarwal, Kakade, Lee, and Mahajan]{agarwal_theory_2021}
Alekh Agarwal, Sham~M. Kakade, Jason~D. Lee, and Gaurav Mahajan.
\newblock On the theory of policy gradient methods: optimality, approximation, and distribution shift.
\newblock \emph{The Journal of Machine Learning Research}, 22\penalty0 (1):\penalty0 98:4431--98:4506, January 2021.
\newblock ISSN 1532-4435.

\bibitem[Arulkumaran et~al.(2017)Arulkumaran, Deisenroth, Brundage, and Bharath]{arulkumaran_brief_2017}
Kai Arulkumaran, Marc~Peter Deisenroth, Miles Brundage, and Anil~Anthony Bharath.
\newblock A {Brief} {Survey} of {Deep} {Reinforcement} {Learning}.
\newblock \emph{IEEE Signal Processing Magazine}, 34\penalty0 (6):\penalty0 26--38, November 2017.
\newblock ISSN 1053-5888.
\newblock \doi{10.1109/MSP.2017.2743240}.
\newblock URL \url{http://arxiv.org/abs/1708.05866}.
\newblock arXiv:1708.05866 [cs, stat].

\bibitem[Beckey et~al.(2022)Beckey, Cerezo, Sone, and Coles]{beckey_variational_2022}
Jacob~L. Beckey, M.~Cerezo, Akira Sone, and Patrick~J. Coles.
\newblock Variational {Quantum} {Algorithm} for {Estimating} the {Quantum} {Fisher} {Information}.
\newblock \emph{Physical Review Research}, 4\penalty0 (1):\penalty0 013083, February 2022.
\newblock ISSN 2643-1564.
\newblock \doi{10.1103/PhysRevResearch.4.013083}.
\newblock URL \url{http://arxiv.org/abs/2010.10488}.
\newblock arXiv:2010.10488 [physics, physics:quant-ph].

\bibitem[Bergholm et~al.(2022)Bergholm, Izaac, Schuld, Gogolin, Ahmed, Ajith, Alam, Alonso-Linaje, AkashNarayanan, Asadi, Arrazola, Azad, Banning, Blank, Bromley, Cordier, Ceroni, Delgado, Di~Matteo, Dusko, Garg, Guala, Hayes, Hill, Ijaz, Isacsson, Ittah, Jahangiri, Jain, Jiang, Khandelwal, Kottmann, Lang, Lee, Loke, Lowe, McKiernan, Meyer, Montañez-Barrera, Moyard, Niu, O'Riordan, Oud, Panigrahi, Park, Polatajko, Quesada, Roberts, Sá, Schoch, Shi, Shu, Sim, Singh, Strandberg, Soni, Száva, Thabet, Vargas-Hernández, Vincent, Vitucci, Weber, Wierichs, Wiersema, Willmann, Wong, Zhang, and Killoran]{bergholm_pennylane_2022}
Ville Bergholm, Josh Izaac, Maria Schuld, Christian Gogolin, Shahnawaz Ahmed, Vishnu Ajith, M.~Sohaib Alam, Guillermo Alonso-Linaje, B.~AkashNarayanan, Ali Asadi, Juan~Miguel Arrazola, Utkarsh Azad, Sam Banning, Carsten Blank, Thomas~R. Bromley, Benjamin~A. Cordier, Jack Ceroni, Alain Delgado, Olivia Di~Matteo, Amintor Dusko, Tanya Garg, Diego Guala, Anthony Hayes, Ryan Hill, Aroosa Ijaz, Theodor Isacsson, David Ittah, Soran Jahangiri, Prateek Jain, Edward Jiang, Ankit Khandelwal, Korbinian Kottmann, Robert~A. Lang, Christina Lee, Thomas Loke, Angus Lowe, Keri McKiernan, Johannes~Jakob Meyer, J.~A. Montañez-Barrera, Romain Moyard, Zeyue Niu, Lee~James O'Riordan, Steven Oud, Ashish Panigrahi, Chae-Yeun Park, Daniel Polatajko, Nicolás Quesada, Chase Roberts, Nahum Sá, Isidor Schoch, Borun Shi, Shuli Shu, Sukin Sim, Arshpreet Singh, Ingrid Strandberg, Jay Soni, Antal Száva, Slimane Thabet, Rodrigo~A. Vargas-Hernández, Trevor Vincent, Nicola Vitucci, Maurice Weber, David Wierichs, Roeland Wiersema, Moritz Willmann, Vincent Wong, Shaoming Zhang, and Nathan Killoran.
\newblock {PennyLane}: {Automatic} differentiation of hybrid quantum-classical computations, July 2022.
\newblock URL \url{http://arxiv.org/abs/1811.04968}.
\newblock arXiv:1811.04968 [physics, physics:quant-ph].

\bibitem[Bhatia(1997)]{bhatia_matrix_1997}
Rajendra Bhatia.
\newblock \emph{Matrix {Analysis}}, volume 169 of \emph{Graduate {Texts} in {Mathematics}}.
\newblock Springer, New York, NY, 1997.
\newblock ISBN 978-1-4612-6857-4 978-1-4612-0653-8.
\newblock \doi{10.1007/978-1-4612-0653-8}.
\newblock URL \url{http://link.springer.com/10.1007/978-1-4612-0653-8}.

\bibitem[Chen et~al.(2020)Chen, Yang, Qi, Chen, Ma, and Goan]{chen_variational_2020}
Samuel Yen-Chi Chen, Chao-Han~Huck Yang, Jun Qi, Pin-Yu Chen, Xiaoli Ma, and Hsi-Sheng Goan.
\newblock Variational quantum circuits for deep reinforcement learning.
\newblock \emph{IEEE Access}, 8:\penalty0 141007--141024, 2020.
\newblock Publisher: IEEE.

\bibitem[Cherrat et~al.(2023)Cherrat, Raj, Kerenidis, Shekhar, Wood, Dee, Chakrabarti, Chen, Herman, Hu, Minssen, Shaydulin, Sun, Yalovetzky, and Pistoia]{cherrat_quantum_2023}
El~Amine Cherrat, Snehal Raj, Iordanis Kerenidis, Abhishek Shekhar, Ben Wood, Jon Dee, Shouvanik Chakrabarti, Richard Chen, Dylan Herman, Shaohan Hu, Pierre Minssen, Ruslan Shaydulin, Yue Sun, Romina Yalovetzky, and Marco Pistoia.
\newblock Quantum {Deep} {Hedging}, March 2023.
\newblock URL \url{http://arxiv.org/abs/2303.16585}.
\newblock arXiv:2303.16585 [quant-ph, q-fin].

\bibitem[Haug and Kim(2021)]{haug_optimal_2021}
Tobias Haug and M.~S. Kim.
\newblock Optimal training of variational quantum algorithms without barren plateaus, June 2021.
\newblock URL \url{http://arxiv.org/abs/2104.14543}.
\newblock arXiv:2104.14543 [quant-ph, stat].

\bibitem[Haug and Kim(2023)]{haug_generalization_2023}
Tobias Haug and M.~S. Kim.
\newblock Generalization with quantum geometry for learning unitaries, March 2023.
\newblock URL \url{http://arxiv.org/abs/2303.13462}.
\newblock arXiv:2303.13462 [quant-ph, stat].

\bibitem[Jerbi et~al.(2021)Jerbi, Gyurik, Marshall, Briegel, and Dunjko]{jerbi_parametrized_2021}
Sofiene Jerbi, Casper Gyurik, Simon Marshall, Hans Briegel, and Vedran Dunjko.
\newblock Parametrized {Quantum} {Policies} for {Reinforcement} {Learning}.
\newblock In \emph{Advances in {Neural} {Information} {Processing} {Systems}}, volume~34, pages 28362--28375. Curran Associates, Inc., 2021.
\newblock URL \url{https://proceedings.neurips.cc/paper/2021/hash/eec96a7f788e88184c0e713456026f3f-Abstract.html}.

\bibitem[Jerbi et~al.(2022)Jerbi, Cornelissen, Ozols, and Dunjko]{jerbi_quantum_2022}
Sofiene Jerbi, Arjan Cornelissen, Māris Ozols, and Vedran Dunjko.
\newblock Quantum policy gradient algorithms, December 2022.
\newblock URL \url{http://arxiv.org/abs/2212.09328}.
\newblock arXiv:2212.09328 [quant-ph, stat].

\bibitem[Kakade(2001)]{kakade_natural_2001}
Sham~M Kakade.
\newblock A {Natural} {Policy} {Gradient}.
\newblock In \emph{Advances in {Neural} {Information} {Processing} {Systems}}, volume~14. MIT Press, 2001.
\newblock URL \url{https://proceedings.neurips.cc/paper_files/paper/2001/hash/4b86abe48d358ecf194c56c69108433e-Abstract.html}.

\bibitem[Meyer(2021)]{meyer_fisher_2021}
Johannes~Jakob Meyer.
\newblock Fisher {Information} in {Noisy} {Intermediate}-{Scale} {Quantum} {Applications}.
\newblock \emph{Quantum}, 5:\penalty0 539, September 2021.
\newblock \doi{10.22331/q-2021-09-09-539}.
\newblock URL \url{https://quantum-journal.org/papers/q-2021-09-09-539/}.
\newblock Publisher: Verein zur Förderung des Open Access Publizierens in den Quantenwissenschaften.

\bibitem[Meyer et~al.(2023{\natexlab{a}})Meyer, Scherer, Plinge, Mutschler, and Hartmann]{meyer_quantum_2023}
Nico Meyer, Daniel~D. Scherer, Axel Plinge, Christopher Mutschler, and Michael~J. Hartmann.
\newblock Quantum {Policy} {Gradient} {Algorithm} with {Optimized} {Action} {Decoding}, May 2023{\natexlab{a}}.
\newblock URL \url{http://arxiv.org/abs/2212.06663}.
\newblock arXiv:2212.06663 [quant-ph].

\bibitem[Meyer et~al.(2023{\natexlab{b}})Meyer, Scherer, Plinge, Mutschler, and Hartmann]{meyer_quantum_2023-1}
Nico Meyer, Daniel~D. Scherer, Axel Plinge, Christopher Mutschler, and Michael~J. Hartmann.
\newblock Quantum {Natural} {Policy} {Gradients}: {Towards} {Sample}-{Efficient} {Reinforcement} {Learning}, August 2023{\natexlab{b}}.
\newblock URL \url{http://arxiv.org/abs/2304.13571}.
\newblock arXiv:2304.13571 [quant-ph].

\bibitem[Niu et~al.(2019)Niu, Boixo, Smelyanskiy, and Neven]{niu_universal_2019}
Murphy~Yuezhen Niu, Sergio Boixo, Vadim~N. Smelyanskiy, and Hartmut Neven.
\newblock Universal quantum control through deep reinforcement learning.
\newblock \emph{npj Quantum Information}, 5\penalty0 (1):\penalty0 1--8, April 2019.
\newblock ISSN 2056-6387.
\newblock \doi{10.1038/s41534-019-0141-3}.
\newblock URL \url{https://www.nature.com/articles/s41534-019-0141-3}.
\newblock Number: 1 Publisher: Nature Publishing Group.

\bibitem[Russell and Norvig(2020)]{russell_artificial_2020}
Stuart Russell and Peter Norvig.
\newblock \emph{Artificial {Intelligence}: {A} {Modern} {Approach} (4th {Edition})}.
\newblock Pearson, 2020.
\newblock ISBN 978-0-13-461099-3.
\newblock URL \url{http://aima.cs.berkeley.edu/}.

\bibitem[Schuld et~al.(2019)Schuld, Bergholm, Gogolin, Izaac, and Killoran]{schuld_evaluating_2019}
Maria Schuld, Ville Bergholm, Christian Gogolin, Josh Izaac, and Nathan Killoran.
\newblock Evaluating analytic gradients on quantum hardware.
\newblock \emph{Physical Review A}, 99\penalty0 (3):\penalty0 032331, March 2019.
\newblock ISSN 2469-9926, 2469-9934.
\newblock \doi{10.1103/PhysRevA.99.032331}.
\newblock URL \url{http://arxiv.org/abs/1811.11184}.
\newblock arXiv:1811.11184 [quant-ph].

\bibitem[Schulman et~al.(2017{\natexlab{a}})Schulman, Levine, Moritz, Jordan, and Abbeel]{schulman_trust_2017}
John Schulman, Sergey Levine, Philipp Moritz, Michael~I. Jordan, and Pieter Abbeel.
\newblock Trust {Region} {Policy} {Optimization}, April 2017{\natexlab{a}}.
\newblock URL \url{http://arxiv.org/abs/1502.05477}.
\newblock arXiv:1502.05477 [cs].

\bibitem[Schulman et~al.(2017{\natexlab{b}})Schulman, Wolski, Dhariwal, Radford, and Klimov]{schulman_proximal_2017}
John Schulman, Filip Wolski, Prafulla Dhariwal, Alec Radford, and Oleg Klimov.
\newblock Proximal {Policy} {Optimization} {Algorithms}, August 2017{\natexlab{b}}.
\newblock URL \url{http://arxiv.org/abs/1707.06347}.
\newblock arXiv:1707.06347 [cs].

\bibitem[Sequeira et~al.(2023)Sequeira, Santos, and Barbosa]{sequeira_policy_2023}
André Sequeira, Luis~Paulo Santos, and Luis~Soares Barbosa.
\newblock Policy gradients using variational quantum circuits.
\newblock \emph{Quantum Machine Intelligence}, 5\penalty0 (1):\penalty0 18, April 2023.
\newblock ISSN 2524-4914.
\newblock \doi{10.1007/s42484-023-00101-8}.
\newblock URL \url{https://doi.org/10.1007/s42484-023-00101-8}.

\bibitem[Skolik et~al.(2022)Skolik, Jerbi, and Dunjko]{skolik_quantum_2022}
Andrea Skolik, Sofiene Jerbi, and Vedran Dunjko.
\newblock Quantum agents in the gym: a variational quantum algorithm for deep q-learning.
\newblock \emph{Quantum}, 6:\penalty0 720, 2022.
\newblock Publisher: Verein zur Förderung des Open Access Publizierens in den Quantenwissenschaften.

\bibitem[Stokes et~al.(2020)Stokes, Izaac, Killoran, and Carleo]{stokes_quantum_2020}
James Stokes, Josh Izaac, Nathan Killoran, and Giuseppe Carleo.
\newblock Quantum {Natural} {Gradient}.
\newblock \emph{Quantum}, 4:\penalty0 269, May 2020.
\newblock ISSN 2521-327X.
\newblock \doi{10.22331/q-2020-05-25-269}.
\newblock URL \url{http://arxiv.org/abs/1909.02108}.
\newblock arXiv:1909.02108 [quant-ph, stat].

\bibitem[Subramanian et~al.(2022)Subramanian, Chitlangia, and Baths]{subramanian_reinforcement_2022}
Ajay Subramanian, Sharad Chitlangia, and Veeky Baths.
\newblock Reinforcement learning and its connections with neuroscience and psychology.
\newblock \emph{Neural Networks}, 145:\penalty0 271--287, January 2022.
\newblock ISSN 0893-6080.
\newblock \doi{10.1016/j.neunet.2021.10.003}.
\newblock URL \url{https://www.sciencedirect.com/science/article/pii/S0893608021003944}.

\bibitem[Sutton and Barto(1998)]{sutton_reinforcement_1998}
Richard~S. Sutton and Andrew~G. Barto.
\newblock \emph{Reinforcement learning - an introduction}.
\newblock Adaptive computation and machine learning. MIT Press, 1998.
\newblock ISBN 978-0-262-19398-6.
\newblock URL \url{https://www.worldcat.org/oclc/37293240}.

\bibitem[Sutton et~al.(1999)Sutton, McAllester, Singh, and Mansour]{sutton_policy_1999}
Richard~S Sutton, David McAllester, Satinder Singh, and Yishay Mansour.
\newblock Policy {Gradient} {Methods} for {Reinforcement} {Learning} with {Function} {Approximation}.
\newblock In \emph{Advances in {Neural} {Information} {Processing} {Systems}}, volume~12. MIT Press, 1999.
\newblock URL \url{https://proceedings.neurips.cc/paper/1999/hash/464d828b85b0bed98e80ade0a5c43b0f-Abstract.html}.

\bibitem[Williams(1992)]{williams_simple_1992}
Ronald~J. Williams.
\newblock Simple statistical gradient-following algorithms for connectionist reinforcement learning.
\newblock \emph{Machine Learning}, 8\penalty0 (3):\penalty0 229--256, May 1992.
\newblock ISSN 1573-0565.
\newblock \doi{10.1007/BF00992696}.
\newblock URL \url{https://doi.org/10.1007/BF00992696}.

\bibitem[Zhan(2002)]{zhan_1_2002}
Xingzhi Zhan.
\newblock 1. {Inequalities} in the {Löwner} {Partial} {Order}.
\newblock In Xingzhi Zhan, editor, \emph{Matrix {Inequalities}}, Lecture {Notes} in {Mathematics}, pages 1--15. Springer, Berlin, Heidelberg, 2002.
\newblock ISBN 978-3-540-45421-2.
\newblock \doi{10.1007/978-3-540-45421-2_1}.
\newblock URL \url{https://doi.org/10.1007/978-3-540-45421-2_1}.

\end{thebibliography}

\newpage
\appendix
\begin{widetext}
\section{Tables for environments description and PQC's}\label{appendix: tables}

    \begin{table*}[!h]
        \centering
        \begin{tabularx}{\textwidth}{|C|C|C|C|C|C|}
        \hline
        \textbf{Environment} & \textbf{State} & \textbf{Action} & \textbf{Reward function} & \textbf{Horizon} & \textbf{Termination criteria} \\
        \hline
        Cartpole & 4 features & 2 actions $A=\{0, 1\}$ & +1 per time step & 200 time steps & Reach horizon or out of bounds \\
        \hline
        Acrobot & 4 features & 3 actions $A=\{0, 1, 2\}$ & -1 + height & 500 time steps & Reach goal or horizon \\
        \hline
        \end{tabularx}
        \caption{Characterization of the environments considered in the numerical experiments. \label{table: envs}}
    \end{table*}
    \begin{table*}[!h]
        \centering
        \begin{tabularx}{\textwidth}{|C|C|C|C|C|}
        \hline
        \textbf{Environment} & \textbf{Policy} & \textbf{Layers} & \textbf{Observables} & \textbf{Batch Size} \\
        \hline
        \multirow{2}{*}{CartPole} & Born & 4 & $\{P_0 , P_1\}$ & 10 \\
        \cline{2-5}
                                    & Softmax     & 4  & $\{P_0 , P_1\}$ & 10 \\
        \hline
        \multirow{2}{*}{Acrobot}  & Born & 5 & $\{P_{0,3} , P_1 , P_2\}$ & 10 \\
        \cline{2-5}
                                    & Softmax     & 5  & $\{P_{0,3} , P_1 , P_2\}$ & 10 \\
        \hline
        \end{tabularx}
        \caption{Characterization of the PQC's considered in the numerical experiments. $P_i$ indicates the projector in the computational basis in decimal. For the Cartpole environment a single-qubit was measured and the probability of each basis state associated to an action. In the Acrobot environment, the action assignmment was made using $int(b)\ \text{mod}\ 3 = a$ for a particular basis state $b$. \label{table: pqcs}}
    \end{table*}

\end{widetext}

\end{document}